# Indication of p + $^{11}$B Reaction in Laser Induced Nanofusion Experiment

N. Kroó[1], L.P. Csernai[2,3,4,*], I. Papp[1,5], M. A. Kedves[1], M. Aladi[1], A. Bonyár[6], M. Szalóki[7,] K. Osvay[8,9], P. Varmazyar[8], and T.S. Biró[1]

(for the NAPLIFE Collaboration)

[1] HUN-REN Wigner Research Centre for Physics, Budapest, Hungary

[2] Dept. of Physics and Technology, University of Bergen, Norway

[3] Frankfurt Institute for Advanced Studies, Frankfurt/M, Germany

[4] Csernai Consult Bergen, Bergen, Norway

[5] HUN-REN Centre for Energy Research, Budapest, Hungary

[6] Budapest University of Technology and Economics, Hungary

[7] University of Debrecen, Debrecen, Hungary

[8] National Laser-Initiated Transmutation Laboratory, University of Szeged, Szeged, Hungary

[9] Department of Optics and Quantum Electronics, University of Szeged, Szeged, Hungary

* laszlo.csernai@uib.no

## ABSTRACT

The NanoPlasmonic Laser Induced Fusion Energy (NAPLIFE) [1] project proposed fusion by regulating the laser light absorption via resonant nanorod antennas implanted into hydrogen rich urethane acrylate methacrylate (UDMA) and triethylene glycol dimethylacrylate (TEGDMA) copolymer targets. In part of the tests, boron-nitride (BN) was added to the polymer. Our experiments with resonant nanoantennas accelerated protons up to 225 keV energy. Some of these protons then led to p + $^{11}$B fusion, indicated by the sharp drop of observed backward proton emission numbers at the 150 keV resonance energy of the reaction. The generation of alpha particles was verified by CR-39 (Columbia Resin #39) nuclear plastic track detectors.

## Introduction

The NanoPlasmonic Laser Induced Fusion Energy, NAPLIFE, [1] project intends to work out affordable technology, which avoids the major obstacles in the leading Inertial Confinement Fusion (ICF) method of Lawrence Livermore National Laboratory (LLNL) National Ignition Facility (NIF) project. Nanotechnology was used earlier in the field of fusion research, in form of long nanowires to avoid plasma reflection of laser light [2-4], The NIF project achieved significant target gain exceeding $Q_t > 1.5$, however, only part of the target fuel was burned, partly due to Rayleigh-Taylor instability and partly faster expansion due to extreme compression than the thermal spreading of fusion reactions. The NAPLIFE project instead aims for simultaneous (time-like [5,6]) ignition with much shorter ignition laser pulse, presently uniquely available at Extreme Light Infrastructure (ELI) - Attosecond Light Pulse Source (ALPS). The simultaneous ignition can only be achieved by regulating the laser light absorption via nanotechnology via **resonant** nanoplasmonic antennas [7]. In this respect the NAPLIFE project is also unique.

Proton-Boron fusion is well discussed and experimented with earlier [8,9], with aneutronic fusion and with non-radioactive fuel, thus we use this version of fusion experiments also.

The present project setup enables non-thermal steps of laser irradiation, absorption, proton acceleration and capture of emitted energy carrying particles, and laser irradiation, fuel feed, and emission of gained energy may be in orthogonal directions making industrial realizations possible.

The properties of localized surface plasmons are being explored, enabling 6 orders of magnitude shorter laser pulses in the femtosecond range, enhancing the local fields in hot spots, leading to screening of positive charges and to ponderomotive acceleration of particles [10–19]. Simultaneous ignition in the whole target volume will be enabled by only two laser beams from opposing directions and by regulating the laser light absorption via resonant nanorod antennas embedded in a thin solid target [1,20,21]. The two-sided irradiation is a simplified and more affordable version of ignition than the similar Double-Cone Ignition [22-25]. In addition, the project plans to economize energy consumption by eliminating thermalization losses in the ignition as the ignition laser beam is monochromatic and linearly polarized. This goal was also formulated earlier [26-36]. This method requires a laser beam pulse duration as short as the light needs to penetrate the target only once [1]. EPOCH (an open-source plasma physics simulation code) PIC (Particle-In-Cell) kinetic model simulations verified that resonant nanorod antennas increase laser light absorption significantly and lead to massive electron resonance in the gold nanorod antennas, while non-resonant antennas do not [20].

Proton acceleration was predicted in PIC kinetic models [21] and observed in laser induced fixed UDMA-TEGDMA copolymer target experiments at the Ti:Sa Hidra laser of Wigner RCP [37]. With the use of the Single Cycle Laser (SYLOS) Experiment Alignment (SEA) at ELI-ALPS [38], the 25 mJ, 12 fs laser pulses accelerated protons up to 1.5 MeV energy from transparent ultra-thin foils. The acceleration mechanism is identified as a hybrid scheme of Coulomb explosion and light sail [39]. Similarly numerous proton acceleration results were reported using Target Normal Sheath Acceleration (TNSA) [40-43].

In the present experiment, the SEA laser pulses have been focused to the target with an f/2 off-axis parabola, Fig. 1. The wave front of the pulses was corrected with a deformable mirror, so that around 60% of the pulse energy was concentrated in the $1/e^2$ focal spot. The orientation of target setup in presented in ref. [39], where the target, Tg1, was used with normal 45-degree from the direction of the laser irradiation, and the Thompson Parabola as well as the CR39 and the proton CCD were measured backwards orthogonally to the target normal.

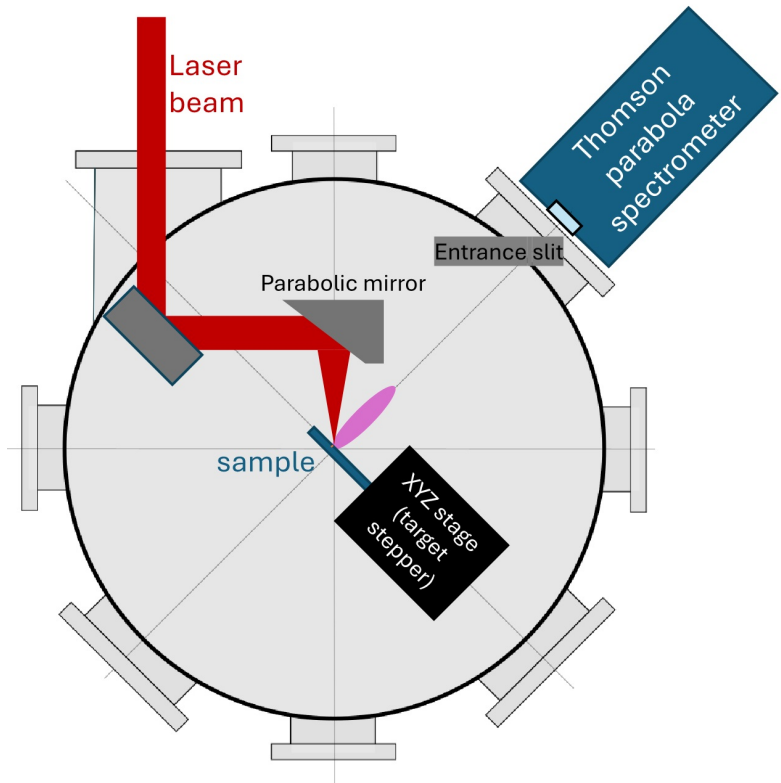

FIG. 1. Schematic of the experimental setup. The flat target plane is orthogonal to the direction of Tomson parabola / laser irradiation. The target is 160 μm thick, so that the laser beam did not penetrate the target but carved out a crater, which emitted a plume. The Thompson parabola detector was positioned at 45 backwards, measuring the emitted charged particles.

The peak intensity in the focus exceeded $10^{18}$ W/cm$^2$. The pulse duration has been varied between 12fs-360fs with the use of an acousto-optical programmable filter (DAZZLER) in the front end of the laser. The focus diameter, ∼ 3 μm and the laser pulse energy, ∼ 25 mJ were kept constant, thus the beam intensity decreased as the irradiation time increased. One sided irradiation was applied with 160 μm thick targets and energetic protons were detected in the backward direction. According to our EPOCH PIC simulation, irradiation of our thick samples with such intensity laser pulses resulted in protons with 225 keV energy [21]. The emitted protons and ions have been measured with a Thomson parabola ion spectrometer, calibrated at low proton energies [44]. The emitted alpha particles have been detected with the use of a CR39 track detector placed in front of the pinhole of the Thomson spectrometer [45].

Our sample, UDMA-TEGDMA copolymer (prepared in 3:1 mass ratio) nanocomposites containing 0.182 m/m% gold nanorods and 2.5 m/m% BN nanoparticles were used for the experiments. The gold nanorods (purchased from *Nanopartz Inc.*) had long and short axes of 85 nm and 25 nm, respectively, with a variance of 10%. The BN nanoparticles (purchased from *Nanografi Co. Inc.*) had a nominal diameter of 65-75 nm. To facilitate binding of the gold nanorods to the surface of the BN particles, their surface was treated with MPTMS (3-mecaptopropyl trimethyloxysilane).

## Results

In our experiment at ELI-ALPS, including boron in the target, at ~ 125 fs laser irradiation length significant drop of the proton number is observed. See Fig. 2. Number of shots
[pulse length (fs), number of shots with BN, number of shots without BN: 12  5 7; 20  - 7; 22  - 7;  25  1 7; 27  - 5; 30  - 6; 40  - 6; 50  6 6; 62  - 6; 73  6 - ; 90  7 - ; 120 11 4; 140  5 3; 170  7 5; 180  6 - ; 220  4 - ; 240  5 - ; 360  4 - ; ]. (For further details of the data contact
Prof. Norbert Kroó, kroo.norbert@wigner.hu)

The total proton signal in this direction is significantly enhanced when using Au nanorod doped targets compared to Au free targets. This enhancement is a consequence of the acceleration of electrons by the plasmonic effect in the resonant nanorod antennas. In these rods large bunches of electrons resonate between the two ends, with the frequency of laser irradiation. These then pull behind themselves the surrounding protons with the Laser Wake Field Acceleration (LWFA) mechanism [21].

The observed protons are in the 100 keV range, with a maximum at 225 keV according to our EPOCH PIC simulation [21]. There are indications that these can lead to a small quantity of deuterium production, (i.e. to nuclear reactions) [37]. This indication is supported by the increase of created crater volumes in case of Au2 nanorods by near to an order of magnitude [46].

Proton Boron (pB) Fusion: To avoid harming radioactivity and dealing with radioactive materials the most practical choice is the aneutronic $p + {}^{11}B$ fusion. This leads to the production of three alpha particles. That reaction proceeds through different channels [47-49], the low branching ratio of the direct ${}^{12}C$ breakup and the sequential decays via the ground state and the first excited state of ${}^{8}Be$. With Au2 doped targets the LWFA process accelerates the protons further in each period of the laser irradiation wave. Thus, increasing irradiation time results in increased proton energy [21], up to ~ 225 keV in present experiments. This increase persists until the increased amount of the resonating electron bunch persists. As some electrons may exit from the nanorods the acceleration decreases at longer times. This reflects the finite lifetime of plasmonic waves. The increasing laser irradiation time leads to increasing maximal proton energy (cf. Fig. 3).

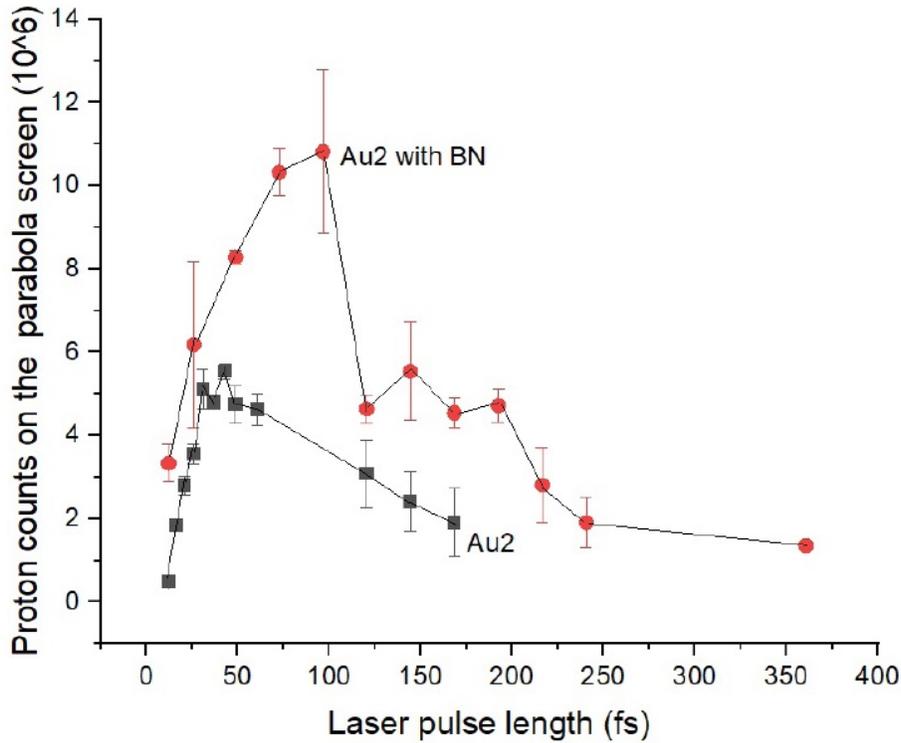

FIG. 2. The integral number of proton signals in a backward direction, measured at ELI-ALPS SEA laser with pulse energy 25 mJ, applying various pulse durations from 12.3 to 360 fs (upper curve with red circles). The maximum beam intensity at 12.3 fs was I= $8.3\times10^{18}$ W/cm$^2$. The target was an UDMA-TEGDMA copolymer with embedded resonant gold nanorod antennas at the density Au2=0.182 m/m%, and boron-nitride (BN) with 2.5m/m% density. This BN number density corresponds to 43% of the number of UDMA-TEGDMA monomers. The experiment where there was no BN in the sample is presented (lower black squares). Neither the gradual increase nor the sudden drop of proton counts is observed. The numbers of shots at different irradiation laser pulse durations were different, changing between 2-12 shots. This resulted in varying error bars.

According to theoretical quantum tunneling calculations, the high density of electrons may reduce the Coulomb barrier of fusion reactions [50-53]. To check aneutronic reactions, a substantial amount boron-nitride was added to the targets. The cross section of the above pB reaction shows a sharp peak at the proton energy $E_R$ = 150 keV [47-49]. This resonance width is about 25 keV and it peaks by one order of magnitude, from 10 to 100 mb. Protons were accelerated via the LWFA mechanism, where electrons resonate along the nanorod with a period of 2.65 fs and attract the protons to move with the same frequency. Each period the protons gain additional energy. In simulations [21] the maximal proton energy exceeds 120 keV soon, in 40-80 fs at an intensity of I =$4\cdot10^{17}$ W/cm$^2$. The acceleration is increased in these simulations to over 150 keV energy with increasing pulse durations [21].

In the present experiment, at the laser pulse duration of 80-90 fs, this maximal energy reaches 120 keV, while at 150-220 fs, it levels at 150 keV in the backward direction (cf. Fig. 3). At the 250 fs pulse duration, the peak experimental intensity was also around I =4 · $10^{17}$ W/cm². The 150 keV proton energy reaches the threshold of the

$$p + {}^{11}B \rightarrow 3\alpha \qquad (1)$$

reaction and depletes the number of protons left in the backward emitted plasma above this threshold. Fig. 3 shows the drop from 11 to 4 million in the proton pixel signal, featuring approximately 70% loss between 100 – 180 fs pulse durations. The drop at 120 fs happens only in targets containing BN. Without BN in the target, the proton count increases less, and there is no sign of any drop at the same pulse duration or energy, see Figure 2. In the BN target Nitrogen may contribute to increased proton content of the plasma.

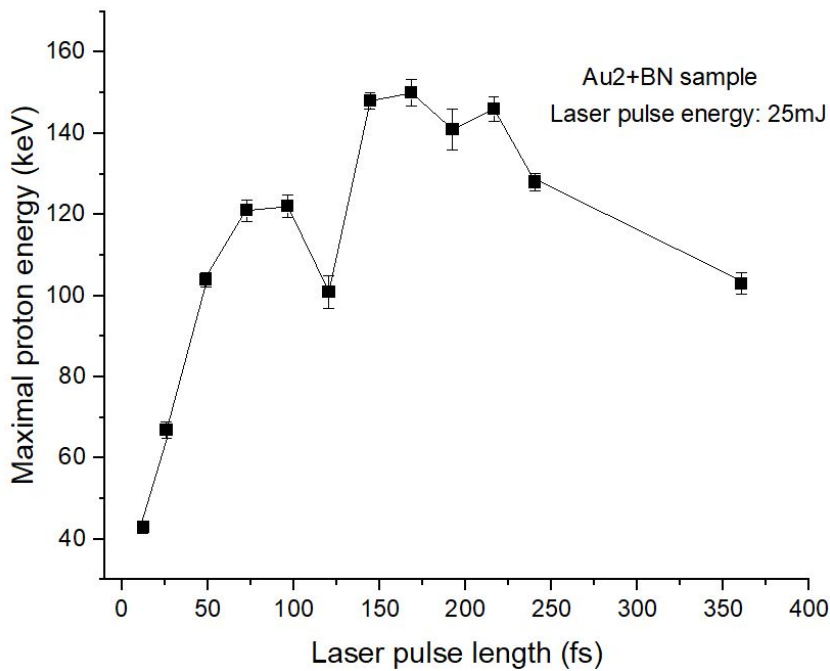

FIG. 3. Maximum proton energy detected in the backward 45-degree direction with respect to the laser beam for different laser pulse durations by Thomson parabola. As can be seen in Fig. 2 at 100-125 fs, the resonant protons are absorbed by the boron in the fusion reactions, and only lower than 150 keV energy protons remain. In the range 150-250 fs pulse durations protons above the 150 keV resonance energy are observed. These originate from those protons, which were already exceedingly well the resonance energy of the p + $^{11}$B cross section.

This clearly indicates that reaction (1) took place in the UDMA-TEGDMA target seeded with BN molecules. The Thomson parabola measurements on the maximal proton energy also support this finding with a rise after the same pulse durations, cf. Fig. 3.

The proton energies were measured at the direction emitted from the impact crater backwards. At this angle the observed proton energy of the emission is under the $E_R$ resonance energy.

In addition, CR-39 nuclear track detector tests were carried out with BN in the target. The histograms for such cases indicate clearly that α particles (wider tracks) were detected in the targets, see Fig. 4.

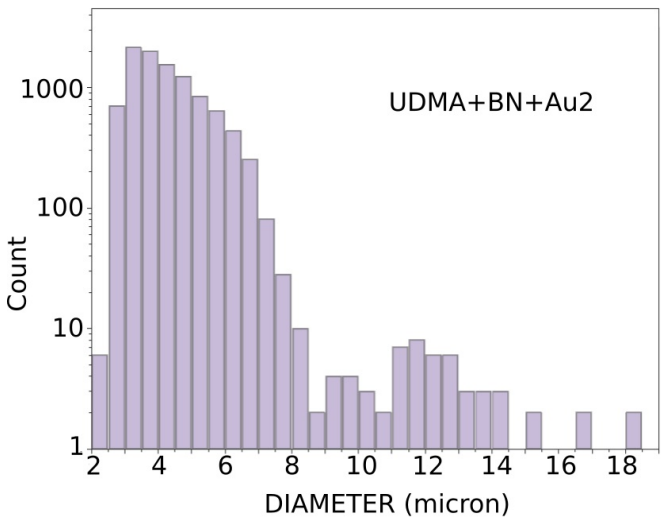

FIG. 4. Number of impact traces versus the diameter of the trace spot in µm in CR-39 detections with boron-nitride in the target shows a second peak at diameter ~12 µm, corresponding to the emitted α particles. The data were obtained from irradiation durations shorter than the proton density drop and at intensity
I = 8.28e18 W/cm$^2$

## Summary

In summary, during these initial validation experiments nanotechnology was assessed: the use of resonant nanorod antennas, with one-sided irradiation of thick fusion targets. Energy production was observed earlier by a significant increase in crater volume, excavated in targets with implanted resonant nanoantennas [46]. Furthermore, resonant nanoantennas in the target have resulted in deuterium production [37]. These findings indicated proton acceleration by the nanoantennas [54]. Subsequently we tested if aneutronic p +$^{11}$B fusion reactions are also facilitated by the accelerated protons. For this purpose, we used a target consisting of the copolymer UDMA-TEGDMA and BN molecules in similar molecular numbers. In these experiments the number of accelerated protons drops significantly at the resonant energy of $E_R$ = 150 keV in the p +$^{11}$B cross section. Furthermore, in C39 emulsion we observe the produced impact traces of α particles.

These observations verify that a small but significant number of fusion reactions have taken place. The future outlook for the project is anticipating further progress when i) the laser pulse energy is increased from 25 mJ to 2 – 20 J at ELI-ALPS, ii) two-sided irradiation is realized to achieve simultaneous volume ignition of flat target, iii) nano-rod antennas will be implanted in the direction of laser beam polarization and in optimized antenna array, iv) the fusion target material will be made more proton and $^{11}$Boron rich.

As mentioned in the introduction the non-thermal setup of the fusion arrangement was tested and larger scale realizations with minimizing losses are promising and were verified in these experimental tests.

## Author contributions

The original idea and evaluation of the plasmonic enhancement of laser pulse energy absorption was performed by N.K. and L.P.Cs. Theoretical PIC model simulations were performed by I.P. Organization and project management of the tasks were done by T.S.B. Laser shooting experiments and Thomson parabola measurements at ELI-ALPS Szeged were performed by M.A., M. A.K, V.P. and K.O. Special target preparation was executed by A.B. and M.S. All authors were involved in manuscript preparation and editing.

## Acknowledgments

Enlightening discussions with J. Rafelski are gratefully acknowledged, experimental help is thanked the ELI-ALPS user support team. We are grateful for their help to J. Csontos and A. Börzsönyi at the ELI-ALPS Single Cycle Laser Group. Contribution to the beam alignment and experimental setup is acknowledged for T. Gilinger and J. Razzaq from NLTL, University of Szeged. We thank all the researchers inside the NAPLIFE project for their respective contributions to the success of this program: M. Csete, D. Vass, A. Szenes, E. Tóth, G. Galbács and his research group at the Szeged University, A. Borók at the Technical University Budapest and our further involved former and recent colleagues at the Wigner RCP, N. Abdulameer, A. Nagyné Szokol, J. Kámán, A. Kumari, M. Veres, R. Holomb, I. Rigó, A. Malik, B. Ráczkevi, A. Inger, K. Zhukovsky and I. Benabdelghani.

L.P. Cs. acknowledges support from Wigner Research Center for Physics, Budapest (2022-2.1.1-NL- 2022-00002). The authors, T.S.B., N.K., I.P., A.B., M.S., M.A. and M. A.K. acknowledge support by the National Research, Development and Innovation Office (NKFIH) of Hungary through the projects 2022-2.1.1-NL-2022-00002, 2018-1.2.1-NKP-201800012 and 2020-2.1.1-ED-2024-0314. The NLTL at University of Szeged has been also supported by NKFIH through the National Laboratory Program, NKFIH-476-4/2021.

## Data availability:

The datasets used and/or analyzed during the current study are available from Prof. Norbert Kroó, [kroo.norbert@wigner.hu](mailto:kroo.norbert@wigner.hu) on reasonable request.",